\RequirePackage[2018-12-01]{latexrelease}
\documentclass{gGAF2e}
\usepackage{xcolor}
\usepackage{graphicx}	
\usepackage{amsmath}
\usepackage{url}
\DeclareMathOperator\erfc{erfc}

\begin{document}

\jvol{00} \jnum{00} \jyear{2012} 

\newcommand{\p}{\partial}
\newcommand{\ie}
{\textit{i}.\textit{e}.\hspace{1mm}}

\markboth{Deepayan Banik and Kristen Menou}{Meridional Circulation Streamlined}


\title{{\textit{Meridional Circulation Streamlined}}}

\author{Deepayan Banik${^\ast}$\thanks{$^\ast$Corresponding author. Email: deepayan.banik@mail.utoronto.ca
\vspace{6pt}} and Kristen Menou${^\ast \dag \ddag}$\\\vspace{6pt}  ${^\ast}$Department of Physics, University of Toronto,
60 St George Street, Toronto, Ontario, M5S 1A7, Canada\\
${\dag}$  Physics \& Astrophysics Group, Dept.   of  Physical  \&  Environmental  Sciences,  University  of  Toronto  Scarborough, 1265 Military Trail, Toronto, Ontario, M1C 1A4, Canada \\
${\ddag}$ David A. Dunlap Department  of Astronomy \& Astrophysics, University of Toronto.
50 St.  George Street, Toronto, Ontario, M5S 3H4, Canada}

\maketitle

\begin{abstract}
Time-dependent meridional circulation and differential rotation in radiative zones are central open issues in stellar evolution theory. We streamline this challenging problem using the ‘downward control principle’ of atmospheric science, under a geostrophic f-plane approximation. We recover the known stellar physics result that the steady-state meridional circulation decays on the length scale proportional to $N/f \times \sqrt{Pr}$, assuming molecular viscosity is the dominant drag mechanism. Prior to steady-state, the meridional circulation and the zonal wind (= differential rotation) spread together via radiative diffusion, under thermal wind balance. The corresponding (4th-order) hyperdiffusion process is reasonably well approximated by regular (2nd-order) diffusion on scales of order a pressure scale-height.  We derive an inhomogeneous diffusion (equiv. advection-diffusion) equation for the zonal flow which admits closed-form time-dependent solutions in a finite depth domain, allowing for rapid prototyping of differential rotation profiles. In the weak drag limit, we find that the time to rotational steady-state can be longer than the Eddington-Sweet time and be instead determined by the longer drag time. Unless strong enough drag operates, the internal rotation of main-sequence stars may thus never reach steady-state. Streamlined meridional circulation solutions provide leading-order internal rotation profiles for studying the role of fluid/MHD instabilities (or waves) in redistributing angular momentum in the radiative zones of stars. Despite clear geometrical limitations and simplifying assumptions, one might expect our thin-layer geostrophic approach to offer qualitatively useful results to understand deep meridional circulation in stars.



\end{abstract}

\section{Introduction}\label{sec:intro}
Meridional circulation \citep{hanasoge2022surface} in stars is important for several reasons, including solar activity cycles and dynamo models \citep{choudhuri1995solar, spruit2002dynamo, weiss2009solar, charbonneau2020dynamo}, latitudinal distribution of sunspots \citep{nandy2002explaining}, advection of angular momentum \citep{miesch2000coupling}, \textit{tachocline} dynamics \citep{spiegel1992solar,rempel2005solar,garaud2008penetration,garaud2009penetration} and the radial transport of chemical species \citep{elliott1999calibration}. Even then, it remains poorly understood and crudely treated in state-of-the-art stellar evolution codes such as a MESA \citep{paxton2010modules}.

Helioseismology \citep{christensen2002helioseismology} shows that the outer convective envelope of the Sun(0.7 - 1 $\rm{R}_\odot$), is rotating differentially \citep{schou1998helioseismic}, while the radiative zone (0.2 - 0.7 $\rm{R}_\odot$) is in approximate solid body rotation \citep{brown1989inferring, goode1991we}. The differential rotation in the convective zone is attributed to different processes like varying Reynolds stresses arising from turbulent convection \citep{rudiger2022differential} or torques induced by solar winds \citep{modisette1967solar,weber1967angular}. Therefore, the radiative zone may be conceived as being torqued by the convection zone from above, leading to the generation of meridional circulation via. a mechanism called \textit{gyroscopic pumping} \citep{hughes2007solar,garaud2009penetration,garaud2010gyroscopic}. 

 It turns out that there is a closely analogous problem in atmospheric science: the \textit{downward control principle} \citep{haynes1991downward} proposed to describe the effect of torques due to waves that are dissipated in the upper atmosphere (stratosphere) of the Earth. As the name suggests, torques on the low-density upper layers can exert `control' on the otherwise inert and dense deeper layers. The process can be understood as follows: alternating \textit{zonal} flows pumped in the upper layer get deflected by Coriolis forces along the \textit{longitudinal} direction similar to Ekman flows \citep{liu2021driving} in the ocean. This results in meridionally periodic regions of mass convergence and divergence at the forced level, thereby generating barotropic pressure gradients on all constant-height surfaces ``throughout the depth of the fluid'' \citep{showman2006deep}. The pressure gradient causes a return flow in the unforced deeper layers, completing the meridional circulation loop. This return flow, in turn, is again deflected by Coriolis forces, generating a deep zonal flow in the same direction as the top-forced wind. Thus, meridional circulation and zonal flow (differential rotation) penetrate jointly through the depth of the rotating stratified fluid.

Meridional circulation of Sun-like stars has seen several analytical \citep{eddington1925circulating,sweet1950importance,spiegel1992solar, garaud2008penetration} and numerical \citep{clement1993hydrodynamical,talon2003finite, garaud2009penetration} treatments under the Boussinesq \citep{rieutord2006dynamics} or the Anelastic \citep{garaud2002rotationally} approximations inspired from geophysical fluid dynamics. Despite these efforts, the process of penetration arguably lacks a simple, intuitive explanation in the stellar astrophysics literature. Using the aforementioned analogy from atmospheric sciences, we re-consider the problem of meridional circulation in stars. We argue that the greater physical intuition gained from the atmospheric science perspective outweighs its geometric/simplifying assumptions, making way for a better qualitative understanding of the process. 

{\color{black} While most previous studies discuss the steady-state meridional circulation \citep{garaud2008penetration,garaud2009penetration}, some that focus on the unsteady problem \citep{elliott1997aspects, wood2012transport, wood2018self} resort to numerical simulations making them computationally expensive. In this work, a first set of analytical solutions is presented for the time-dependent problem, allowing for convenient reproducibility. Such a solution is important as stars are low Prandtl number environments and hence may not reach rotational steady-state in their main-sequence lifetimes. In the context of stellar evolution codes like MESA, the time-dependent analytical solution will serve to accurately transport angular momentum and chemicals through meridional circulation. Additionally, differential rotation profiles obtained may be probed for magnetic/fluid instabilities/waves, thereby allowing the calculation of appropriate diffusion coefficients that parameterize transport beyond what is currently implemented in stellar models.}

We adopt the primitive $f-$plane meteorological framework of \cite{showman2006deep}, which investigates the depth of penetration of zonal jets and their associated meridional circulation into the molecular envelope of Jupiter. In \S\ref{sec:2}, we discuss the assumptions and derive an expression for the steady-state meridional streamfuction, thereby confirming that it decays on a length scale {\color{black}proportional to $N/f \times \sqrt{Pr}$, a product of buoyancy and rotation frequency ratio and the Prandtl number} \citep{spiegel1992solar,garaud2002rotationally}. In \S\ref{sec:3}, we derive a single 1D advection-diffusion equation for the zonal velocity in log-pressure coordinates and explain both steady-state and time-dependent analytical solutions, comparing them with Dedalus2 numerical solutions in \S\ref{sec:4}. Finally, we conclude in \S\ref{sec:5}. 

\section{The meridional streamfunction} \label{sec:2}

Figure \ref{fig:0}a shows the internal structure of a solar-type star with an outer convective zone, an inner radiative zone, and the core at the center.  The zonal, meridional, and vertical coordinates are $\phi, \theta$, and $z$, respectively. {\color{black} The convective zone applies a zonal torque on the underlying radiative zone represented as prograde and retrograde winds in figure \ref{fig:0}b.}

\subsection{Assumptions}

\begin{figure*}
    \centering
    \includegraphics[width=0.9\textwidth]{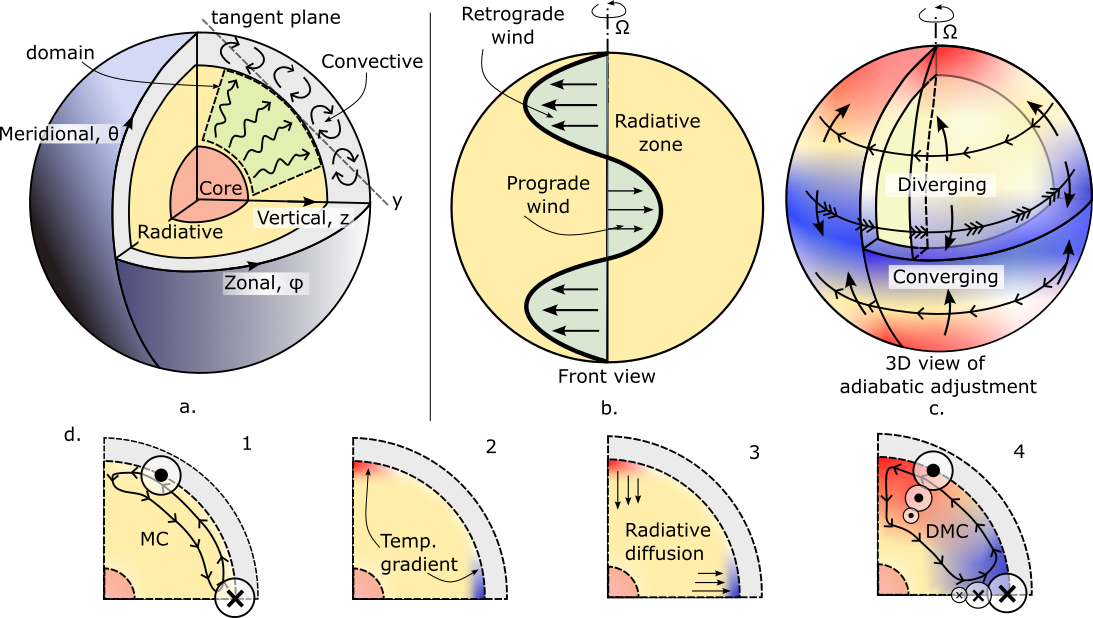}
    \caption{{\color{black}a. The interior structure of a solar-type star showing the convection zone, the radiative zone, the core, the principle coordinates $(z,y, \phi)$, and the tangent plane. b. The structure of zonal forcing/torque applied by prograde and retrograde winds at the top of the radiative zone. c. Converging and diverging longitudinal flows due to the action of Coriolis force on the top-forced wind. d. The numbered panels, in order, show the various elements of the meridional circulation. The prograde and retrograde zonal winds are represented by arrows pointing out of and into the plane. The red and blue regions represent the baroclinic structure that results from the zonal torque according to the thermal wind balance. The initial meridional circulation (MC in panel d1) cell is shown as a loop with arrows representing the direction of circulation. The final circulation cell is deep (DMC in panel b4) and corresponds to a smoother vertical gradient of zonal velocity. Note that the schematics do not include any representation for vertical stratification, but it is present in general.}}
    \label{fig:0}
\end{figure*}

We use the linearized primitive equations of meteorology in pressure coordinates \citep{holton1973introduction} to model the radiative zone. The details are elucidated in \cite{showman2006deep}. Our assumptions are as follows:

\begin{itemize}
\item The flow is zonally symmetric \ie quantities do not vary with the azimuthal angle (or longitude), $\phi$.
        \item f-plane (see figure \ref{fig:0}a): We use Cartesian coordinates rather than spherical. The f-plane approximation uses a fixed value of the Coriolis parameter $f(=2 \varOmega \sin \theta)$ set to correspond to mid-latitudes ($\theta=\theta_0$). Here $\varOmega$ is the background rotation rate.
        \item The horizontal extent $L_y$, is close to the entire length from the pole to the equator. This stretches the f-plane approximation beyond the formal limit but captures the proper horizontal scale of interest.
        \item The pressure scale height $H$ is assumed to be constant, as would happen in an isothermal radiative zone. It is also stably stratified with a Brunt frequency $N$.
        \item A constant gravitational acceleration $g$ is adopted; for the sun, $g$ changes by a factor no more than 2 over most of the radiative zone.
        \item Hydrostatic balance: this is a common assumption in both stellar astrophysics and geophysical fluid dynamics (GFD). It implies that vertical motion with velocity $w$ is significantly suppressed compared to horizontal velocities $u$ and $v$, which is a direct consequence of the aspect ratio of the flow $\lambda$ being $\ll 1$. We note that a deep radiative zone formally violates the thin layer approximation built in our formalism.                
        \item {\color{black}Geostrophic balance: The Rossby number of the flow is small in the stellar radiative zone i.e. $Ro \sim \Delta\varOmega/\varOmega\sim 0.1$ \citep{garaud2009penetration}}. Here, $\Delta\varOmega$ is the deviation of the latitudinal differential rotation from the background $\varOmega$. Inertial forces are therefore dominated by rotation. {\color{black} Note that other forcings like gravitational contraction are stronger sources of $\Delta \varOmega$ \citep{gouhier2021axisymmetric} leading to larger Rossby numbers, but we do not consider them here.}
        \item The hydrostatic balance and geostrophic balance combine into the thermal wind balance. This implies that a radial gradient in velocity corresponds to a horizontal gradient in temperature.
        \item Diffusion of momentum and temperature are approximated by  linear relaxations \citep{showman2006deep}; also see appendix \ref{sec:hyperdiff}.
        \item The feedback of the radiative zone on surrounding convection zones is ignored.
    \end{itemize}

\subsection{The steady-state streamfunction}

The independent variables representing the vertical and the meridional coordinates are pressure $p$ and distance $y$, respectively. Time is denoted by $t$. The zonal momentum, thermal wind, mass continuity and thermodynamic energy equations for a stratified hydrostatic atmosphere with momentum forcing \citep{showman2006deep} are given by, respectively,
\begin{equation}
    \frac{\p u}{\p t}-f v=\frac{u_F-u}{\tau_F}
    \label{eq:1}
\end{equation}
\begin{equation}
    f p\frac{\p u}{\p p}=R\frac{\p T'}{\p y}
    \label{eq:2}
\end{equation}
\begin{equation}
    \frac{\p v}{\p y}+\frac{\p w}{\p p} = 0
    \label{eq:3}
\end{equation}
\begin{equation}
    \frac{\p T'}{\p t} - \frac{N^2 H^2}{R p} w = \frac{T'_E-T'}{\tau_R}
    \label{eq:4}
\end{equation}
where $u$, $v$, and $w(\equiv \p p/\p t)$ are the zonal, meridional, and vertical wind speeds in the rotating frame of reference, $f(=2 \varOmega \sin \theta)$ is the Coriolis frequency parameter, $\theta$ is the latitude and $R$ is the ideal gas constant. The momentum forcing relaxes $u$ towards $u_F$, and the thermal forcing relaxes $T$ towards $T'_E$. The corresponding frictional drag ($\tau_F$) and radiative ($\tau_R$) relaxation times approximate the usual operators for viscous and heat diffusion. The Brunt frequency $N$ for the stratified radiative zone with a temperature distribution $T=T_0(p)+T'$, where  $T'$ is a small perturbation from the reference state $T_0$, is given by
\begin{equation}
    N=\sqrt{\frac{g}{H}\left(\frac{R}{C_P}-\frac{p}{T}\frac{\text{d} T}{\text{d} p}\right)},
    \label{eq:5}
\end{equation}
where $C_P$ is the specific heat of the constituent gas at constant pressure or along isobars, $H=RT/g$ is the isothermal scale height, and $g$ is the acceleration due to gravity. The nonlinear advection terms in the momentum equation can be ignored in first approximation owing to the low Rossby number $Ro$ of the flow resulting from geostrophic balance \citep{vallis2017atmospheric}. We thus have four equations (\ref{eq:1})-(\ref{eq:4}) and four variables $u$, $v$, $w$ and $T'$.


In this work, we consider thermal forcing to be zero \ie $T'_E=0$ \citep{busse1981eddington} to focus on the role of momentum forcing for meridional circulation. On eliminating $u$ and $T'$ from (\ref{eq:1})-(\ref{eq:4}) and substituting $v$ and $w$ with their corresponding expressions in terms of the meridional streamfunction $\psi$ ($v=-\p\psi/\p p$ and $w=\p \psi/\p y$), {\color{black}in steady-state we obtain}
\begin{equation}
    \frac{p^2}{f \tau_R}\frac{\p u_F}{\p p} = \frac{p^2}{\tau_R}\frac{\p}{\p p}\left(\tau_F\frac{\p \psi}{\p p}\right) + \frac{N^2 H^2}{f^2}\frac{\p^2 \psi}{\p y^2}.
    \label{eq:6}
\end{equation}

The domain of interest is divided into two parts following the setup of \cite{showman2006deep}: the top layer corresponds to the near-convection zone where momentum forcing is applied (RHS of equation \ref{eq:1}), while the deeper, stably stratified layer below is passive and only driven by the circulation in the top forced layer. The forcing velocity $u_F$, relaxation timescales $\tau_R$ and $\tau_F$, and the Rossby deformation radius $L_D(=NH/f)$ are piecewise constant along the vertical coordinate $p$, demarcating the two layers described. {\color{black} It is to be noted that the full two-layer setup will not be entirely solved in this work. In the following sections, the role of the upper layer is to simply inform the boundary condition on the bottom layer. However, the functional form of $u_F$ is mentioned here for completeness.
\begin{equation}
    \displaystyle u_F(p,y) = 
\begin{cases} 
    U_F \cos(ly), \text{if} \quad p \leq p_{rcb}\text{ i.e. in top layer } \\ 
    0, \quad \quad \quad \quad \text{if} \quad p > p_{rcb}\text{ i.e. in bottom layer} 
\end{cases} 
\end{equation}
where, $U_F$ is the velocity forcing magnitude and $p_{rcb}$ is the pressure at the radiative-convective boundary.} In the passive layer, where $u_F=0$, equation (\ref{eq:6}) reduces to,
\begin{equation}
    L_D^2\frac{\p^2 \psi}{\p y^2} + p^2\frac{\tau_F}{\tau_R}\frac{\p^2 \psi}{\p p^2} = 0.
    \label{eq:7}
\end{equation}

This equation, together with appropriate boundary conditions, gives the steady-state meridional streamfunction in a rotating hydrostatically and geostrophically balanced fluid such as the radiative zone of a Sun-like star. We see that $\psi$ admits a separable solution of the form $\psi(p,y)=\psi_p(p)\psi_y(y)$, with $\psi_y(y)=\cos(ly)$ \ie periodic in latitude, {\color{black}where $l (=2\pi/L_y)$} can represent the latitudinal wavenumber of the shallow jets forced in the top layer {\color{black}extending from the equator to the poles}. Substituting the above in equation (\ref{eq:7}) we obtain an ordinary differential equation for the vertical component of the streamfunction $\psi_p$ as
\begin{equation}
   p^2 \frac{\p^2 \psi_p}{\p p^2} = \bar{\varOmega}^2 \psi_p
    \label{eq:8}
\end{equation}
where $\bar{\varOmega}^2=L_D^2 l^2 \tau_R/\tau_F
$. It admits the polynomial solution for $\psi$:
\begin{equation}
    \psi(p,y)=\left[A p^{n_1} + B p^{n_2}\right]\cos(ly)
    \label{eq:9}
\end{equation}
where, $n_1$ and $n_2$ are roots of the equation $n(n-1)=\bar{\varOmega}^2$. Constants $A$ and $B$ may be determined from interface and boundary conditions. The vertical {\color{black}dependence} of the steady-state streamfunction, which determines the steady-state depth of the meridional circulation, is therefore strictly dependent on $\bar{\varOmega}$ which is the product of the Burger number (Bu = $L_D^2 l^2$) and the Prandtl number (Pr = $\tau_R/\tau_F$) as has already been established in literature e.g. \citet{spiegel1992solar,garaud2002rotationally,gilman2004limits,garaud2008penetration}. {\color{black}For the sake of a more popularly used expression, we may rewrite $\bar{\varOmega}=(N/f)\sqrt{Pr}Hl$, explicitly showing the ratio of buoyancy and rotation frequencies.} We now use this setup and formalism to explore the time-dependent behavior of the meridional circulation in a top-forced radiative zone.

\section{Time-dependent zonal velocity equation}
\label{sec:3}


We use the same equations (\ref{eq:1})-(\ref{eq:4}) to derive a single equation for the time-dependent penetration of the zonal wind. In the absence of thermal forcing, $T'_E=0$, and assuming the variables $u$, $v$, $w$ and $T'$ to be meridionally periodic with a wavenumber $l$, so that
\begin{equation}
    u=\bar{u}\cos(ly), v=\bar{v}\cos(ly), w=\bar{w}\sin(ly) \hspace{1mm} \text{and} \hspace{1mm} T'=\bar{T'}\sin(ly),
    \label{eq:10}
\end{equation}
where the barred quantities are functions of $t$ and $p$, we obtain the following from equations (\ref{eq:1})-(\ref{eq:3}) in the passive zone ($u_F=0$) below the momentum-driven layer,
\begin{equation}
    \frac{\p \bar{u}}{\p t}-f \bar{v}=\frac{-\bar{u}}{\tau_F}
    \label{eq:11}
\end{equation}
\begin{equation}
    f\frac{\p \bar{u}}{\p \ln p} = l R\bar{T'}
    \label{eq:12}
\end{equation}
\begin{equation}
    l\bar{v}=\frac{\p \bar{w}}{\p p}
    \label{eq:13}
\end{equation}
Additionally, assuming a rapid thermal adjustment ($\p T'/\p t = 0$) leading to a balance of entropy advection and thermal diffusion, equation (\ref{eq:4}) gives,
\begin{equation}
    \bar{w} = \frac{R\bar{T'}p}{N^2 H^2 \tau_R}.
    \label{eq:14}
\end{equation}
Combining equations (\ref{eq:11})-(\ref{eq:14}) gives us a 1D second-order inhomogeneous diffusion equation for the zonal wind in pressure coordinates:
\begin{equation}
    \frac{\p \bar{u}}{\p t} + \frac{\bar{u}}{\tau_F} = \frac{\p }{\p p}\left(\frac{p^2}{L_d^2 l^2 \tau_R}\frac{\p \bar{u}}{\p p} \right) = \left(\frac{\p k}{\p z} + k\right) \frac{\p \bar{u}}{\p z} + k\frac{\p^2 \bar{u}}{\p z^2} 
    \label{eq:15}
\end{equation}
where, $L_D=(NH)/f$ is the Rossby deformation radius, the log-pressure vertical coordinate $z=\ln p$  and $k=1/(L_D^2 l^2 \tau_R)$. It can be rewritten as a linear advection-diffusion equation for the evolution of the zonal wind (RHS of equation \ref{eq:15}). Accounting for the simplification of the radiative diffusion operator as Newtonian linear relaxation, equation (\ref{eq:15}) shares close similarity with the hyperdiffusion equation obtained for the radial profile of differential rotation by \cite{spiegel1992solar}. In its most general form, equations (\ref{eq:15}) may have $L_D, \tau_R, \tau_F$, and $k$ that dependent on $p$. However, it is possible to find useful approximate analytical solutions to equation (\ref{eq:15}) if the above parameters are constant. We apply the Dirichlet boundary condition  $\bar{u}(0,t)=u_F$ at the top of the passive domain as an idealized form of zonal forcing. {\color{black} The bottom of the domain, $L$, has a free slip (no shear) boundary, and hence a Neumann boundary condition given by $u_z(L,t)=0$.} Under these conditions, one can use solution C3 in \cite{van1982analytical} for the steady-state,
\begin{multline}
    \bar{u}(z)=u_F \left[\exp{\left(\frac{j+k}{2k}z\right)}+\frac{j+k}{j-k}\exp{\left(-\frac{j-k}{2k}z+\frac{jL}{k}\right)}\right]/ C(z);\\ j=\sqrt{k^2+\frac{4k}{\tau_F}}; C(z)=1+\frac{j+k}{j-k}\exp{\left(\frac{jL}{k}\right).} \label{eq:16}
\end{multline}
An approximate time-dependent solution for the initial condition $\bar{u}(z,0)=0$ can be obtained from C7 in \cite{van1982analytical} and rewritten as
\begin{multline}
        \bar{u}(z,t) = \frac{u_F B(z,t)}{2C(z)};m^{\pm}=\exp{\left(-\frac{k \pm j}{2k}\right)}; n^{\pm}=\frac{z\pm jt}{2\sqrt{+kt}};\\
 B(z,t) = m^- \erfc(n^-)+m^+ \erfc(n^+) + \exp{\left(\frac{jL}{k}\right)} \times\\ \left[\frac{j+k}{j-k}m^+ \erfc{\left(\frac{L}{\sqrt{+kt}}-n^+\right)} + \frac{j-k}{j+k}m^- \erfc{\left(\frac{L}{\sqrt{+kt}}-n^-\right)}\right]\\-
 \frac{k\tau_F}{2}\exp{\left(-L-\frac{t}{\tau_F}\right)}\erfc{\left(\frac{2L-z-kt}{2\sqrt{+kt}}\right)}
 \label{eq:17}
\end{multline}
where `$\erfc$' is the complementary error function. We comment on the limit of validity of this solution in the following section.

\section{Applications}\label{sec:4}


Let us consider a radiative zone that is initially under solid body rotation, $\varOmega$. At time $t=0$, a zonal wind, $u_F$, at the outer boundary ($z=0$) starts torquing this stably stratified layer. A fast, purely dynamical, adiabatic/geostrophic adjustment {\color{black}moves around fluid parcels creating local convergence and divergence at the surface (see figure \ref{fig:0}c)}, thereby setting up a meridional circulation (refer \S \ref{sec:intro}), the depth of which depends on the ratio of rotation and stratification $\varOmega/N$, only. The circulation advects the background stratification to establish a meridional temperature gradient. Section 3 of \cite{spiegel1992solar} gives a detailed description of the process.
 

Our solution aims to model the subsequent much slower evolution. The meridional temperature gradient developed in the adjusted layer (see Figure \ref{fig:0}d panel 2) also impacts the vertical temperature profiles. Following a rapid thermal adjustment occurring on a short thermal timescale \citep{spiegel1992solar}, radiative diffusion propagates the horizontal temperature inhomogeneities further down; Figure \ref{fig:0}d panel 3. As a result, the zonal wind and the meridional circulation spread deeper into the radiative envelope (Section 4 \cite{spiegel1992solar}). The Rayleigh frictional drag (or viscosity) opposes this by acting on the vertical shear of the zonal wind. In steady-state, the meridional circulation penetrates to a depth until these two effects balance each other out. This slower process depends on the Prandtl number in addition to rotation and stratification, as discussed in \S \ref{sec:2} for the steady-state meridional streamfunction.

\subsection{Steady-state zonal velocity}

Two main parameters influence the steady-state solution, $k$ and $\tau_F$, as is evident from equation (\ref{eq:15}). The depth of penetration of zonal velocity and, thus, the meridional circulation crucially depends on both of them. Other parameters that also affect the nature of the solution are the boundary conditions \citep{garaud2008penetration}, namely, $u_F$ and free shear. 

\begin{figure}
    \centering
    \includegraphics[width=0.55
    \textwidth]{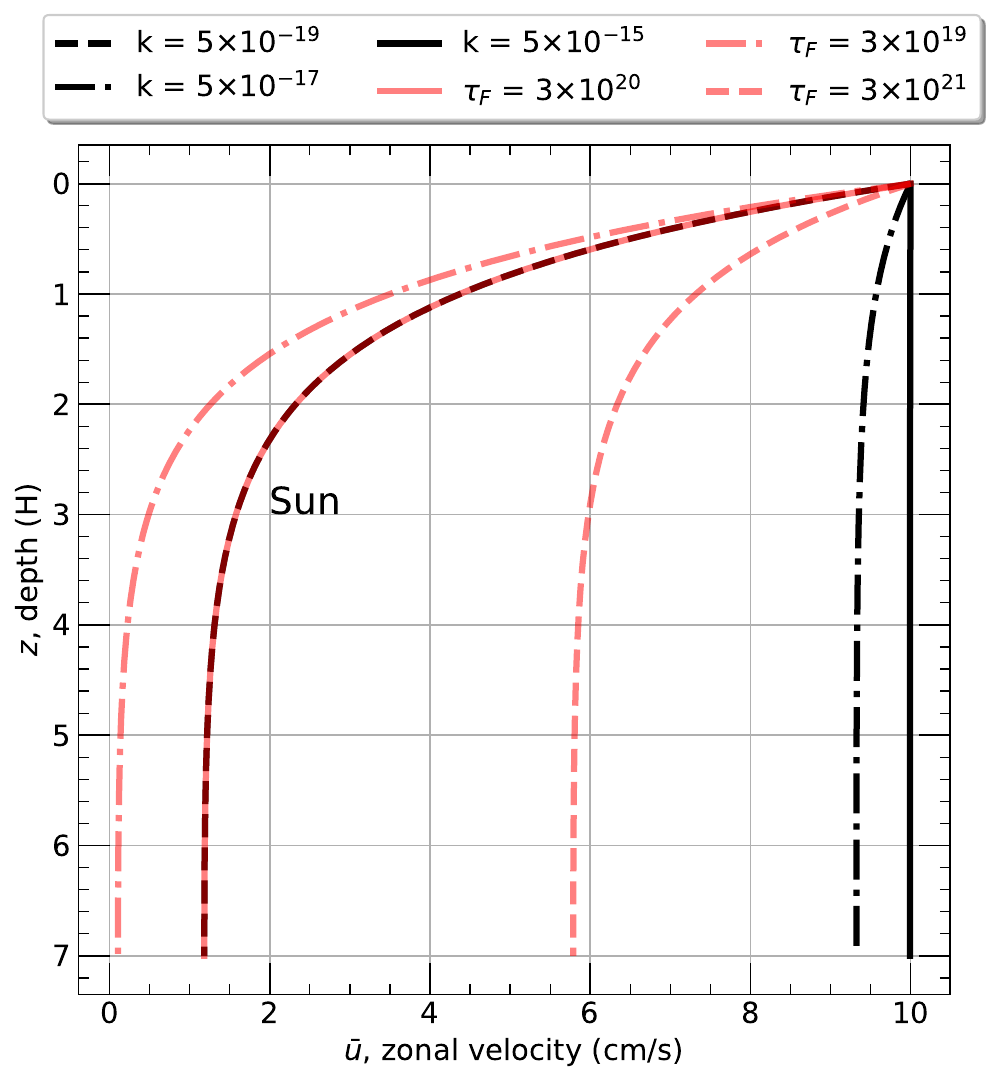}
    \caption{Analytical solution for steady-state zonal wind varying with the vertical coordinate indicating the depth of penetration of the meridional circulation. The top and bottom of the radiative zone are indicated by $z=0$ and $z=7$, respectively. Different values of $k$ (black) include the effects of rotation $f$, stratification $N$, meridional wavenumber $l$, scale height $H$ and the radiative relaxation timescale $\tau_R$ according to $k=1/(L_D^2 l^2 \tau_R)$. See \S \ref{sec:3} for details. The effect of the viscous drag timescale $\tau_F$, on the depth of penetration is shown by red lines. The black dashed line overlapping the red solid line, on the left, depicts the steady-state zonal velocity profile in the solar radiative zone.}
    \label{fig:1}
\end{figure}

Here, $k$ accounts for the radiative timescale, rotation, and stratification via $\tau_R$, $f$, and $N$, respectively. Figure \ref{fig:1} shows vertically varying steady-state zonal wind profiles for different values of $k$ (black, constant $\tau_F=10^{13}\hspace{1mm}\rm{yrs} = 3\times10^{20} \rm{s}$) and $\tau_F$ (red, constant $k=5\times10^{-19}\hspace{1mm} {\rm{cm}}^2/{\rm{s}}$) using expression (\ref{eq:16}). These are values adopted as representative of the Sun \citep{menou2004local, menou2006magnetorotational}, the zonal velocity profile for which is labeled in the figure  (black dashed, red solid lines). The radiative-convective interface is at $z=0$, and the bottom of the domain is at $z=\ ln (10^{3})=7$, considering the passive radiative zone to extend through pressures varying three orders in magnitude. The forcing zonal velocity, $u_F$, is $10\hspace{1mm}cm/s$ without loss of generality. The bottom boundary shears freely. Thus, there is no hard constraint on the velocity, allowing it to assume any value.

Under strong radiative diffusion \ie, small $\tau_R$, the zonal flow, and the corresponding meridional circulation penetrate deep. Greater penetration is also achieved, albeit during adiabatic adjustment, for a weak stratification or strong rotation following the expression of $k$. Figure \ref{fig:1} also shows the effect of the frictional drag timescale $\tau_F$ on the penetration depth. A small $\tau_F$ corresponds to a stronger drag, limiting the spread from radiative diffusion early so that the zonal velocity forcing cannot penetrate deep, other parameters remaining the same.


{\color{black}From the same figure, the zonal wind is seen to penetrate all the way through the radiative zone of the Sun at \textit{steady-state}. This is in reasonable agreement with the \textit{steady-state} estimate made using equation (5.19) of \citet{spiegel1992solar} by substituting \textit{our} value of the Prandtl number. Their formulation includes the role of horizontal turbulence rather than frictional drag. However, a value for the Prandtl is not specified and is left to general interpretations. \citet{spiegel1992solar} also have some estimates for the Sun's \textit{current} degree of penetration in the absence of drag, which will be discussed in the next section.}

\subsection{Time-dependent penetration}


{\color{black} Previous time-dependent studies include \citet{elliott1997aspects}, which numerically solves a set of equations close to \citet{spiegel1992solar} in spherical coordinates, although reporting results only till one solar age; \citet{wood2012transport,wood2018self} perform full-blown three-dimensional numerical simulations of the radiative-convective interface including gravity waves, convective penetration and magnetic fields. They claim that the leading order dynamics are well captured by `balanced' models and may not require complex simulations. This sets the stage for the results obtained from our model, which, as discussed below, provide a significantly cheaper alternative and succinct explanation.}

\begin{figure}
    \centering
    \includegraphics[width=0.9
    \textwidth]{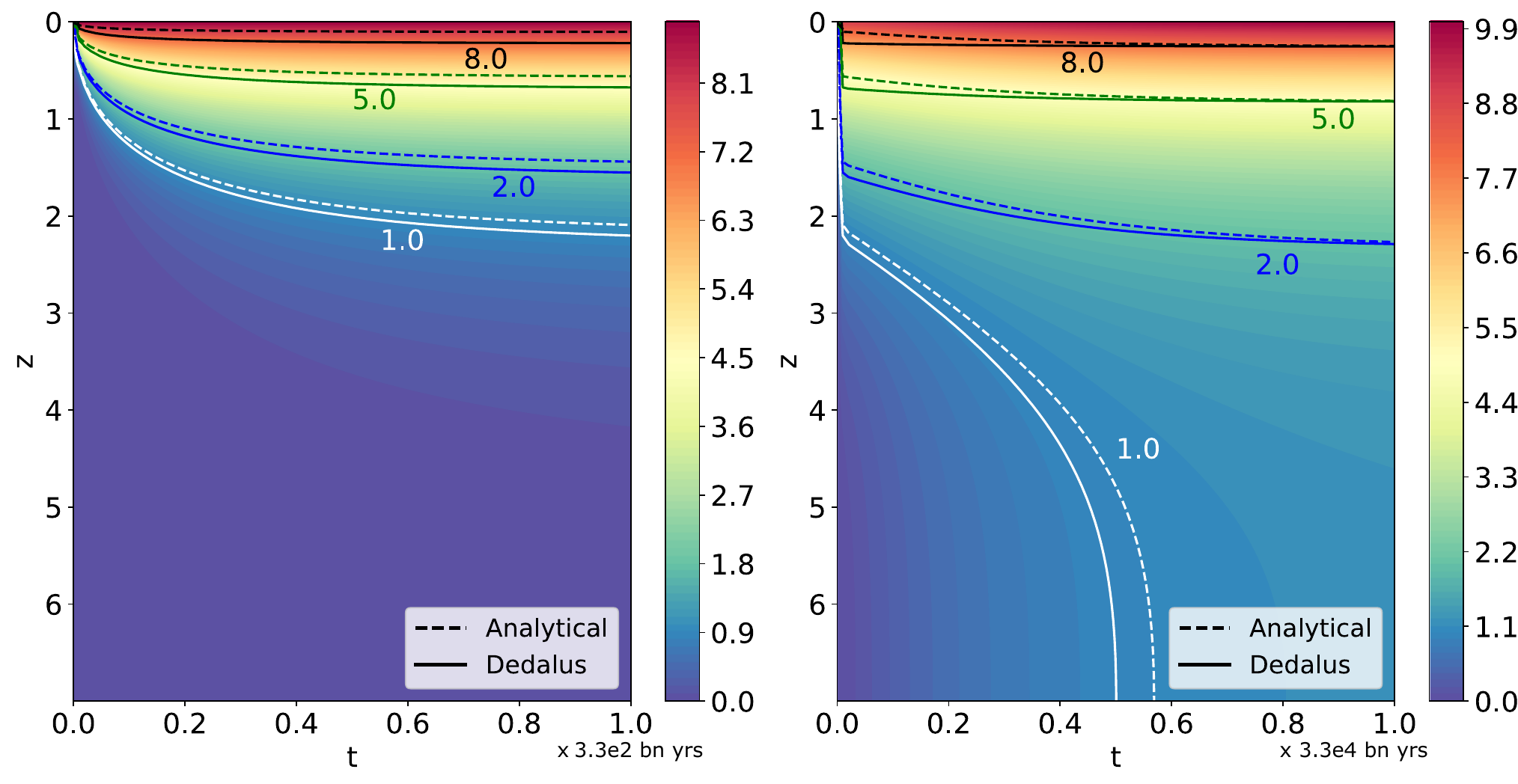}
    \caption{Time-dependent analytical and numerical solutions for the evolution of zonal velocity contours at \textit{early} stages (till $t = 10^{19}s$). The zonal wind forcing $u_F$ at the radiative-convective interface ($z=0$) is $10 cm/s$. The boundary condition at the bottom is free shear \ie $\p u/\p z = 0$. The background colormap represents the magnitude of the zonal velocity, as obtained from the analytical solution, at different depths on the ordinate as time proceeds on the abscissa. Specific contours of the numerical Dedalus2 solution is shown using solid lines, while the dashed lines represent the analytical solution in addition to the background colormap. The left and right panels show early and late stages of penetration, respectively. Note: Maximum value of the zonal velocity on the colorbar is NOT 10 despite the imposed BC, especially earlier (left panel). This is a limitation of the analytical solution.}
    \label{fig:2}
\end{figure}

{\color{black}We use Dedalus2 \citep{burns2020dedalus} to numerically solve equation (\ref{eq:15}) and support the analytical solution. Dedalus is a pseudo-spectral code used to solve initial, boundary, or eigen-value problems with the flexibility of specifying the equations symbolically. A Chebyshev basis is used to discretize our 1D domain and Dirichlet and Neumann boundary conditions are applied at the top ($\bar{u}(0,t)=u_F$) and bottom ($u_z(L,t)=0$), respectively. Post discretization the system is evolved through the SBDF2 timestepper.}  Figure \ref{fig:2} compares the time-dependent analytical solutions (cf. expression (\ref{eq:17})) with numerical solutions. The temporal evolution of the zonal velocity is represented as contours on a spacetime plot at \textit{early} and \textit{late} stages, respectively. The value of $k=5\times10^{-19}\hspace{1mm}{\rm{cm}}^2/{\rm{s}}$ and  $\tau_F=10^{13}\hspace{1mm}\rm{yrs} =3\times10^{20}\rm{s}$. The dashed and solid lines represent the analytical solution (also the background colormap) and Dedalus, respectively. The different intensities of the zonal wind clearly shows the penetration of the meridional circulation. 

Two distinct evolutionary phases are observed: one \textit{fast} and the other \textit{slow}. {\color{black}If we assume rotation, stratification, and horizontal and vertical length scales to be constants, the aforementioned phases can be explained by two competing effects determined by the magnitudes of $\tau_R$ and $\tau_F$.} A strong radiative diffusion renders relatively fast penetration. The zonal velocity diffuses to a certain depth in time $\sim\tau_R$ while obeying the thermal wind balance. The Rayleigh drag, acting weakly, develops gradually and slows the penetration down until it comes to a complete halt on the timescale $\sim$ $\tau_F$. At steady-state radiative diffusion and Rayleigh drag balance each other out. This distinctive evolution is clear in the weak drag limit, $\tau_R < \tau_F$, as shown in figure \ref{fig:2}. Thus, the time to steady-state is dictated by the longer drag time in this case. The internal rotation profile may therefore never reach steady-state if the drag is too weak, making the time-dependent solution directly relevant. {\color{black}However, in the limit of stronger drag \ie $\tau_R \gtrsim \tau_F$, there is simply one evolutionary phase with hardly any penetration (not shown here; can be reproduced with the Dedalus script shared at the end).}

{\color{black}Figure \ref{fig:4} (right panel) shows the time evolution of zonal velocity in the depths of the Sun, which, under all simplifying assumptions of the current work, must reach \textit{steady-state} in $3\times10^{20}\rm{s}$ ($10^{13}\hspace{1mm}\rm{yrs}$) or approximately 2000 times its \textit{current} age. {\color{black} This is greater than both the Eddington-Sweet time and nuclear time for the Sun, which have values of $2.2\times 10 ^{11}\rm{yrs}$\citep{spiegel1992solar} and $10^{10}\rm{yrs}$, respectively.} Hence, the Sun should run out of hydrogen fuel well before reaching rotational steady-state unless stronger drag is present. The left panel in Figure \ref{fig:4} compares the current and steady-state zonal velocity profiles using our formalism. The steady-state zonal/meridional flow penetrates all the way to the bottom of the radiative zone as also pointed out in Figure \ref{fig:1}. In the current Sun, the tachocline extends through 0.75 pressure scale heights to a depth located at $0.625\rm{R}_\odot$ from the center of the Sun.

\citet{spiegel1992solar} present some current-Sun estimates of penetration of differential rotation under various assumptions like zero drag and thin tachocline, amounting to distances $\sim$ 0.314$\rm{R}_\odot$ and 0.44$\rm{R}_\odot$ from the center of the Sun for similarity and numerical solutions, respectively. These suggest a much deeper penetration than the shallow, helioseismically observed value at 0.68$\rm{R}_\odot$ \citep{elliott1999calibration}. In order to truncate the penetration, horizontal turbulence is invoked, which in the strong limit ensures that steady-state is reached early enough for theory to match observations. Thus, rotational steady-state is not achieved in its main-sequence lifetime lest additional effects such as drag or anisotropic turbulence are invoked.

On the other hand, the parameter $N/f \sqrt{Pr}$ (say $\beta$) is not a constant for the Sun as shown in Figure 10 of \citet{garaud2009penetration}. Their work suggests that tachocline depth should be restricted to where $\beta \lesssim 1$ \ie, the region where the Taylor-Proudman regime is not broken. This happens at about 0.55$\rm{R}_\odot$ from the center of the Sun. For comparison, our idealized Sun-model considers a constant value of $\beta=0.89$, which implies a weak Taylor-Proudman constraint, allowing penetration through the entire depth at steady-state. {\color{black}Here, we stick to a constant value for $\beta$ so that an analytical solution is possible. However, one might solve for domain-dependent $\beta$ numerically (see Appendix \ref{sec:raddiff}) by implementing non-constant coefficients $k$ and $\tau_F$.}


\begin{figure}
    \centering
    \includegraphics[width=0.55
    \textwidth]{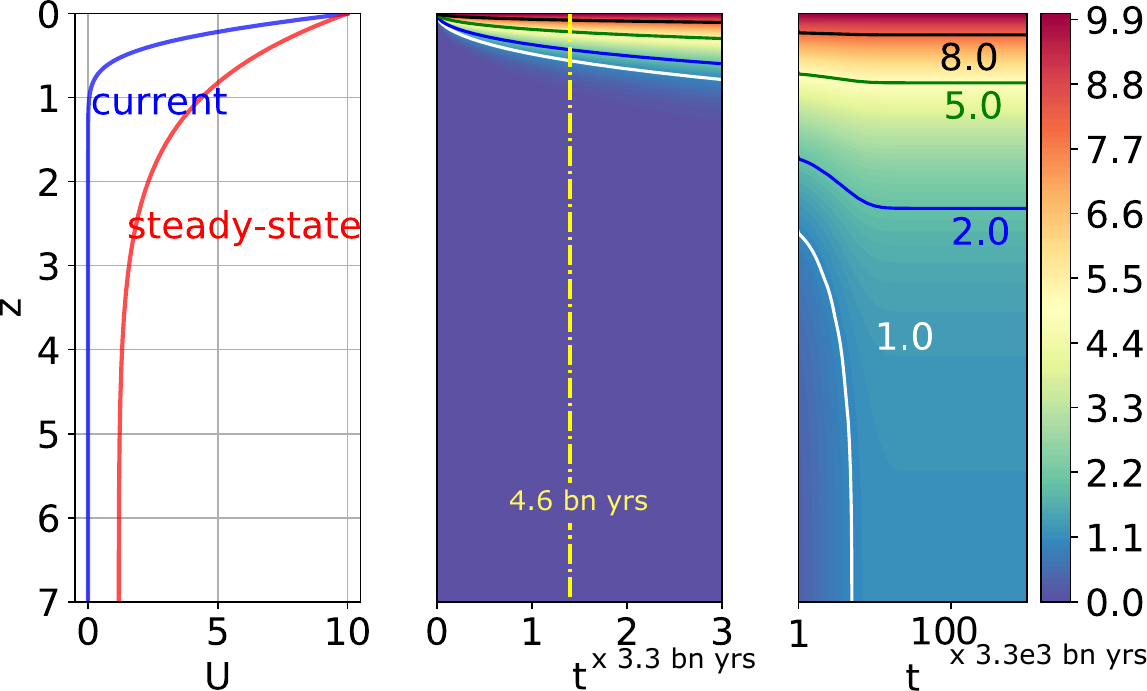}
    \caption{{\color{black}Left panel. Current ($4.6\times10^{9}$ $\rm{yrs}$, solid blue) and steady-state (solid red) zonal velocity profiles varying with depth in the radiative zone of the Sun. Middle and right panels. Time evolution of the penetration of zonal wind in the Sun. The dash-dotted yellow vertical line in the middle panel corresponds to the current solar age, while the extreme right edge of the right panel corresponds to steady-state.}}
    \label{fig:4}
\end{figure}

It is to be noted that the accuracy of the analytical solution is parameter dependent and breaks down at the boundaries because the $u=0$ initial condition throughout the domain cannot match the $u=u_F$ boundary condition. The discrepancy is more apparent at earlier times than late (see maximum colorbar values in figure \ref{fig:2}). The analytical solution also breaks down when $k$ and $1/\tau_F$ differ from one another by more than two orders of magnitude. Also, the linearized Rayleigh drag does not capture the scale dependence of an ordinary 2nd-order differential operator. Similarly, the radiative diffusion should ideally be replaced by a 4th-order hyperdiffusion term as in \cite{spiegel1992solar}. Despite these hefty caveats, the GFD-inspired thin-layer $f$-plane solution captures the leading order penetration phenomena with significant accuracy. It can be used to prototype differential rotation profiles in stellar interiors quickly, thereby allowing probes into the onset of magneto/hydrodynamic instabilities(waves) that potentially transport angular momentum in radiative envelopes of main-sequence stars. However, shortcomings of the analytical solution, variable parameters and hyperdiffusion call for detailed numerical solutions. 

\section{Conclusions}\label{sec:5}

We employ the ``downward control principle'' from atmospheric science to address the time-dependent penetration of meridional circulation and differential rotation in the radiative zones of Sun-like stars due to zonal momentum forcing from the upper convective zone. Assuming geostrophic balance in an f-plane and adopting pressure as the vertical coordinate, we gain novel insights through a relatively simplified approach.

Our steady-state meridional streamfunction reaffirms the established understanding in stellar physics that the circulation decays over a length scale proportional to $N/f \times \sqrt{Pr}$. Rotation and radiative diffusion are drivers of deep penetration, while a stable stratification and viscous drag act to oppose the effect. The differential rotation, which is a combination of the zonal wind and meridional circulation, propagates deep following a fast adiabatic adjustment to the thermal wind balance. We derive a time-dependent, one-dimensional, inhomogeneous diffusion equation which reasonably approximates the fourth-order hyperdiffusion process \citep{spiegel1992solar} on a scale comparable to the pressure scale height. Closed-form analytical solutions allow rapid prototyping of differential rotation profiles at any time during the evolution of a main-sequence star.  

\citet{spiegel1992solar} have previously built a steady-state model for the penetration of differential rotation by invoking a strong turbulent (horizontal-only) viscosity to match helioseismic profiles. Using our time-dependent formalism, we also find that a sufficiently strong drag is required to truncate the penetration and reach steady-state, otherwise, the internal rotation of main-sequence stars may keep evolving through their lifetime. {\color{black}This could have potential consequences for chemical mixing and transport of angular momentum through meridional circulation and hydrodynamic/hydromagnetic instabilities in stellar interiors \citep{busse1981eddington}.}

Extensions of this work include the study of the effect of mean-molecular weight gradients on the penetration of meridional circulation and the use of computed differential rotation profiles to probe the onset of magneto-/hydrodynamic instabilities in stellar radiative zones, in an attempt to better understand the transport of angular momentum in Sun-like stars.

\section*{Acknowledgements}
DB is supported by the Department of Physics at University of Toronto. KM is supported by the National Science and Engineering Research Council of Canada. The authors would like to thank the anonymous reviewers for their reviews which significantly enhanced the paper.

\section*{Data Availability}

The data underlying this article is available at: \url{https://github.com/Textydeep/Meridional-Circulation-Dedalus2.git}





\bibliographystyle{gGAF}
\bibliography{gGAFguide}


\appendix

\section{Hyperdiffusion} \label{sec:hyperdiff}

In this appendix, we make a further connection to the hyperdiffusion solutions of \citet{sakurai1969spin,spiegel1992solar}. Replacing the linear relaxation for temperature perturbations by the usual 2nd order heat diffusion operator in equation (\ref{eq:4}) and separating out the horizontal coordinate $y$, leaves us with the following replacement to equation (\ref{eq:14}),
\begin{equation}
    \Bar{w}=-\left(\frac{\kappa}{H^2}\right)\left(\frac{Rp}{N^2H^2}\right)\frac{\p^2 \Bar{T}}{\p z^2}.
    \label{eq:b1}
\end{equation}
Here, $\kappa$ is the coefficient of radiative heat diffusion. Combining equations (\ref{eq:11}-\ref{eq:13}) and (\ref{eq:b1}) we get
\begin{equation}
    \frac{\p \bar{u}}{\p t} + \frac{\bar{u}}{\tau_F} = \frac{\p }{\p p}\left(\frac{-\kappa p}{L_d^2 l^2 H^2}\frac{\p^3 \bar{u}}{\p (\ln p)^3} \right) =  -k \frac{\p^3 \bar{u}}{\p z^3} - k\frac{\p^4 \bar{u}}{\p z^4} 
    \label{eq:B2}
\end{equation}

where $k=\kappa/(l L_D H)^2$, is the same as $k=1/(L_D^2 l^2 \tau_R)$ mentioned previously in \S \ref{sec:3}, assuming we set $\tau_R =  H^2/\kappa$. By comparing our hyperdiffusion equation to that in \citet{spiegel1992solar}, we recognize a similar 4th-order diffusive operator. However, the inhomogeneous nature of hyperdiffusion in our equation differs from the homogeneous result in \citet{spiegel1992solar}, possibly because of the thin layer assumption $ (h << H) $ adopted in their work.

Dimensional analysis shows that the regular 2nd-order diffusion evolves on a time scale of order $z^2 \tau_R$. By contrast, the 4th-order hyperdiffusion operator evolves on a time scale of order $z^4 \tau_R$. These two timescales match for $z=1$, i.e., on a scale equal to the pressure scale height $H$, which is also the scale we chose to relate $\tau_R$ and $\kappa$. In other words, we expect a regular diffusion treatment to approximate the actual hyperdiffusion process relatively well on scales of order H. We also note that the rather large scale height $H \sim 0.1\rm{R}_\odot$ in a main sequence star like the Sun implies a modest dynamic range of only $z \sim 1 \text{-}7$ over the full radiative zone.

{\color{black} We also recognize that linear
relaxation is usually applicable to optically thin media, while the radiative stellar interior is optically thick. We argue that linear relaxation, which is a first-order approximation for the regular diffusion operator, may be used so long the perturbation temperature is significantly smaller than the background profile i.e. the temperature does not vary drastically with height in the medium \citep{2008ApJ...682..559S}.}

For all these reasons, we focus on a simpler second-order diffusion treatment in our main discussion to gain insight and obtain leading-order results. We recognize that numerical solutions to the exact hyperdiffusion equation (\ref{eq:B2}) could be obtained relatively easily and would be necessary when the parameters in the hyperdiffusion equation vary with coordinates $z$. Nevertheless, we expect the main understanding gained from our second-order diffusion analysis with constant parameters to remain broadly valid and useful.

\section{Vertically varying radiative diffusion} \label{sec:raddiff}

{\color{black} We use Dedalus2 to solve equation (\ref{eq:15}) with $k$ varying in the vertical, which is the more generally relevant scenario. We see the steady-state zonal wind profiles in Figure \ref{fig:B1} (right panel) corresponding to different depths at which radiative diffusion is shut off (left panel). {\color{black}The $k$ profiles are chosen to be nearly non-differentiable so that maximum effect on the solution may be observed.}  That radiative diffusion is the clear driver of deep penetration is evident from the truncation of the spread of zonal wind in the lower region for models with larger Prandtl number. Another feature is the accumulation of kinetic energy in the strongly radiative-diffusive upper region, leading to higher-magnitude winds. The shallower this region of energy accumulation, the stronger the wind.}

\begin{figure}
    \centering
    \includegraphics[width=0.5
    \textwidth]{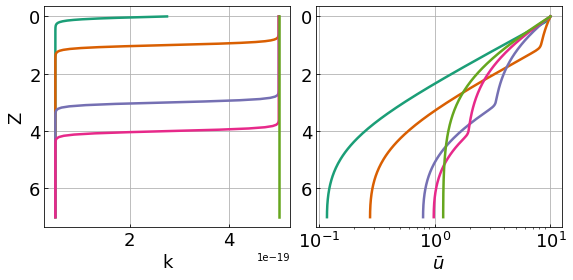}
    \caption{Domain-dependent profiles of $k$ and the corresponding steady-state zonal velocity profiles that result from them. The maximum value of $k$ is the same as that for the Sun, and the minimum value is one-tenth the maximum. The x-axis of the second plot is in log-scale and clearly shows the accumulation of kinetic energy in the radiatively diffuse upper layer. See text for details.}
    \label{fig:B1}
\end{figure}


\end{document}